\documentclass[11pt]{JHEP3}
\usepackage{latexsym}
\usepackage{cite}
\usepackage{epsfig}
\input epsf
\title{Tadpole Analysis of Orientifolded Plane-Waves} 

\author{Aninda Sinha and Nemani V Suryanarayana\\

DAMTP, Centre for Mathematical Sciences, \\ 
Wilberforce Road, Cambridge CB3 0WA, UK \\ 
~~\\
{\rm E-mail}:~\email{
a.sinha,
v.s.nemani@damtp.cam.ac.uk} }

\preprint{
\hepth{0209247}\\
DAMTP-2002-100}

\abstract{We study orientifolds of type IIB string theory in the
plane-wave background supported by null RR 3-form flux $F^{(3)}$. We
describe how to extract the RR tadpoles in the Green-Schwarz formalism
in a general setting. Two models with orientifold groups $\{1, \Omega
\}$ and $\{ 1, \Omega\, I_4\}$, which are T-dual to each other, are
considered. Consistency of these backgrounds requires 32 D9 branes for
the first model and 32 D5 branes for the second one. We study the
spectra and comment on the
heterotic duals of our models.}

\begin{document} 
\section{Introduction}

It has recently emerged that type IIB string theory admits
supersymmetric plane-wave backgrounds \cite{blaua, blaub} supported by
Ramond-Ramond (RR) fluxes. String theory in these backgrounds turns
out to be solvable in the light-cone Green-Schwarz(GS) formalism for
both closed \cite{metsaev, mettsey, russotsey} and open strings
\cite{atishparv, bgmnn, rashmi, yoji, pascal}. Many of these
backgrounds can be obtained as Penrose limits of known ones. This fact
enables understanding of their holographic duals in terms of some
sectors of the CFT dual to the original background \cite{bmn}.

It is well known that one can generate new vacua of string theory by
gauging a subgroup of discrete symmetries. For
instance, the symmetry group of type IIB string theory in the Minkowski
space contains the following discrete subgroup:
\begin{equation}
\label{2bg}
{\cal G} = \{1, \Omega, (-1)^{F_L}, (-1)^{F_R}, (-1)^{F_s}, \Omega
(-1)^{F_L}, \Omega (-1)^{F_R}, \Omega (-1)^{F_s}\}
\end{equation}
where $\Omega$ represents the world-sheet parity operator,
$(-1)^{F_L}$ and $(-1)^{F_R}$ represent the space-time fermion number
operators coming from the world-sheet left movers and the right movers
respectively and $(-1)^{F_s}\equiv (-1)^{F_L + F_R}$. One can consider
various subgroups of ${\cal G}$ and gauge them to obtain new vacua.
Orientifolds are the vacua obtained by gauging a discrete subgroup
which contains at least one element with $\Omega$. A simple example is
the type I theory with $SO(32)$ gauge group obtained as an orientifold
of type IIB by the group $\{1, \Omega\}$.

With the new plane-wave vacua of string theory at hand it is natural
and interesting to study their orientifolds. In \cite{bgmnn} the
authors have considered the orientifold of type IIB theory in the
background of the maximally supersymmetric plane-wave supported by a
null RR 5-form flux \cite{blaua}. More orientifold models have been
considered in \cite{wyllard}. In this paper we would like to explore
orientifolds of type IIB in the background of the supersymmetric
plane-wave
\begin{eqnarray}
\label{ppwave2}
ds^2 &=& dx^+dx^- - f^2x^ix^i (dx^+)^2 
+ dx^idx^i + dx^\alpha dx^\alpha \cr 
&& ~~~ F^{(3)}_{+12} = F^{(3)}_{+34} = -2f.
\end{eqnarray}
where $F^{(3)}$ is the field strength of the RR two form of type
IIB, $i = 1, 2, 3, 4$ are the indices along the plane-wave but
transverse to the light cone directions $(x^+, x^-)$ and $\alpha = 5,
6, 7, 8$ are four flat directions transverse to the plane-wave.

The background (\ref{ppwave2}) can be obtained as a Penrose limit of
$AdS_3\times S^3 \times T^4$ solution of type IIB. The
world-sheet theory of closed strings in this background has been
solved in the Green-Schwarz light-cone formulation in \cite{bmn,
russotsey}. Similarly, open strings in this background have been
quantized in \cite{yoji}.

For constructing an orientifold of this background, we first note that
the world-sheet parity operator $\Omega$ generates a symmetry of this
model because the RR 2--form is invariant under the action of
$\Omega$. Therefore, one could gauge this symmetry combined with some
discrete space-time symmetry and extract the resultant theory. For
example type IIB on (\ref{ppwave2}) admits the following discrete
symmetry groups: (i) $\{ 1, \Omega\}$, (ii) $\{ 1, \Omega I_4\}$ where
$I_4$ is reflection along four of the eight coordinates transverse to
the light-cone direction with the condition that the reflection $I_4$
leaves the RR field invariant, (iii) $\{ 1, \Omega I_8 \}$ where $I_8$
is reflection of all $x^i$ and $x^\alpha$.

In this paper we consider the following two models:
\begin{enumerate}
\label{models}
\item{ Type IIB on PP$_6 \times {\mathbb R}^4$/$\{1, \Omega\}$}
\item{ Type IIB on PP$_6 \times T^4$/$\{1, \Omega I_{5678}\}$}
\end{enumerate}
where we have denoted the space-time of (\ref{ppwave2}) along $(x^+,
x^-, x^1, x^2, x^3, x^4)$ directions by PP$_6$.
 Typically, orientifolds suffer from tadpole divergences. For the
consistency of the theory, one has to ensure that such tadpoles are
absent. For example, type IIB on ${\mathbb R}^{9,1}$/$\{1, \Omega\}$
has a non-vanishing RR 10--form tadpole. This is cancelled by
introducing 32 D9--branes in the vacuum. 
In what follows, we analyze the RR tadpoles for the
models above. 

In section 2, we review the RR tadpole calculation from the point of
view of the NSR formalism . We then describe how to extract the RR
exchange contributions of the Cylinder(C), M\"obius strip(MS) and
Klein Bottle(KB) diagrams using the light-cone Green-Schwarz
formulation.  In section 3, we describe the action of the discrete
symmetries on the states of the Hilbert space and use the results of
section 2 to calculate the RR tadpoles of our models. We find that the
RR tadpole cancellation requires 32 D9-branes in the first model and
32 D5 branes in the second model. One can arrange the spectrum of
`massless' states in representations of the isometry group. We carry
out this analysis for the open string massless fields in detail for
the first model. The plausible Heterotic dual description of our
orientifold vacua are also considered. We conclude in section
4. Appendix A contains a review of the relevant features of closed and
open strings quantization in the plane-wave background. Appendix B
contains some details on the regularizations of the zero point
energies of bosons and fermions in plane-wave backgrounds.

\section{Revisiting the Orientifold} 

String theory in the background of plane-waves of the kind
(\ref{ppwave2}) with RR fields can be solved in the Green-Schwarz
formalism in the light-cone gauge. As such, it is necessary to
establish the rules for tadpole computations in the Green-Schwarz
formalism. Before this, it is useful to briefly review how the RR
tadpole is calculated in the NSR
formalism\cite{sagnotii,gimpol,atishrev,sgnotrev}.

\subsection{RR Tadpole in NSR Formalism}
Recall that there are three diagrams at the one loop order in
perturbative string theory with unoriented closed and open strings:
the Klein bottle (KB), the M\"obius strip (MS) and the Cylinder
(C). The KB amplitude corresponds to the exchange of closed strings
between two orientifold planes. The MS amplitude corresponds to that
between an orientifold plane and a D-brane and C amplitude corresponds
to that between two D-branes. Conformal invariance allows us to view
these as either tree channel or loop channel diagrams. These two
channels are related to each other by modular transformations of the
corresponding diagrams. Let $l$ denote the modular parameter of a
diagram in the tree channel ( a closed string of length $2\pi$
propagating for a time $2\pi l$) and $t$ denote that in the loop
channel (a closed string of length $2\pi$ or an open string of length
$\pi$ propagating for a time $2\pi t$). Then they are related to each
other by the following equations.
\begin{equation}
\label{modtrans}
{\rm KB}: ~~t = {1\over 4l}, ~~~~ 
\hbox{MS}: ~~t = {1\over 8l}, ~~~~
{\rm C}: ~~t = {1\over 2l}. 
\end{equation}
The amplitudes can be expressed in terms of the following standard
modular functions:
\begin{eqnarray}
\label{fdefone}
f_1(q) &=& q^{\frac{1}{12}}\prod_{n=1}^{\infty}(1 - q^{2n})
, ~~~~~~~~\,
f_2(q) = \sqrt{2} q^{\frac{1}{12}}\prod_{n=1}^{\infty}(1 + q^{2n})\cr
f_3(q)&=& q^{-\frac{1}{24}}\prod_{n=1}^{\infty}(1 + q^{2n-1}), ~~~~
f_4(q) = q^{-\frac{1}{24}}\prod_{n=1}^{\infty}(1 - q^{2n-1}).
\end{eqnarray}
These functions satisfy the Jacobi Identity (abstruse identity)
\begin{equation}
\label{abstruse}
f_3^8(q) = f_2^8(q) + f_4^8(q)
\end{equation}
and have the following modular transformation properties
\begin{equation}
\label{fmodprop}
f_1(e^{-\pi /s}) = \sqrt{s}f_1(e^{-\pi s}), ~~ 
f_3(e^{-\pi /s}) =  f_3(e^{-\pi s}), ~~
f_2(e^{-\pi /s}) = f_4(e^{-\pi s}).
\end{equation}
One way to extract the tadpoles is to calculate the diagrams KB, MS
and C in loop channel as traces over the open (for C and MS) or closed
(for KB) string Hilbert space and convert them into tree channel using
relations (\ref{modtrans}). The tadpoles arise in the limit
$t\rightarrow 0$ (equivalently $l\rightarrow \infty$). In a
supersymmetric theory with Bose-Fermi degeneracy at each level, all
three amplitudes KB, MS and C vanish individually. In the loop
channel, the vanishing is because of the Bose-Fermi degeneracy of the
string spectrum. From the tree channel point of view, in the NSR
formalism, they vanish because of the cancellation between the
contributions of NSNS and the RR exchange amplitudes(though each of
NSNS and RR fields give divergent contributions to the amplitude). For
the RR tadpole cancellation, one has to extract the RR field
contribution alone for all the diagrams, add them and require that the
sum vanishes. For this one needs to analyze how the boundary
conditions in tree channel map onto those in the loop channel.

For example, for the orientifold model Type IIB on ${\mathbb
R}^{9,1}/\{1, \Omega\}$ one gets the following contributions to the RR
exchange amplitude: $\frac{v_{10}}{256}\int_0^\infty {dt \over t^6}$
times
\begin{eqnarray}
\label{rrtadone}
{\rm KB}&:~~& 32 \frac{f_4^8(e^{-2\pi t})}{f_1^8(e^{-2\pi t})} \cr
{\rm MS}&:~~& -\frac{f_2^8(e^{-2\pi t}) f_4^8(e^{-2\pi t})}
{f_1^8(e^{-2\pi t}) f_3^8(e^{-2\pi t})}\{ {\rm
Tr}(\gamma_\Omega^{-1} \gamma_\Omega^T)\} \cr
{\rm C}&:~~& \frac{f_4^8(e^{-\pi t})}{f_1^8(e^{-\pi t})} \{ 
({\rm Tr}(\gamma_1))^2\}
\end{eqnarray}
where $v_{10} = V_{10}/(4\pi^2 \alpha')^3$ with $V_{10}$ being the
regularized space-time volume, $\gamma_\Omega$ and $\gamma_1$ are the
matrix representation of the elements $\Omega$ and $1$ respectively on
the open-string Chan-Paton(CP) factors. To extract the tadpole
consistency condition one factorizes this amplitude in the tree
channel using the modular transformations (\ref{modtrans},
\ref{fmodprop}) to get
\begin{equation}
\label{tadpole1}
\frac{v_{10}}{16}\int_0^\infty dl \{ 32^2 
- 64{\rm Tr}(\gamma_\Omega^{-1} \gamma_\Omega^T) 
+ ({\rm Tr}(\gamma_1))^2\}.
\end{equation}
For SO projection on the CP factors we have $\gamma_\Omega^T =
\gamma_\Omega$ and one chooses $\gamma_1 = 1$. Since $\gamma_\Omega$ is
symmetric for the SO projection, by a unitary change of basis one can
make $\gamma_\Omega = 1$. With this, expression in (\ref{tadpole1})
becomes:
\begin{equation}
\frac{v_{10}}{16}\int_0^\infty dl\, (n_9 - 32)^2
\end{equation}
where $n_9 = {\rm Tr}(1)$ is the number of D9-branes. Therefore the RR
tadpole in this model vanishes for $n_9 = 32$ giving rise to the Type
I theory with a gauge group SO(32).

\subsection{RR Tadpole from Light-cone Green-Schwarz String} 

Unlike in the NSR case, in GS formalism there is no such
straightforward way of interpreting the amplitudes in the tree
channel. The aim of this subsection is to describe a method to extract
the RR exchange contributions in the tree channel from the loop
channel amplitudes in the GS formalism.

The basic idea is to use the fact that in the tree channel, we have
contributions from the `NSNS' and `RR' fields exchanged and one can
extract the RR contribution alone by inserting a projection operator
which projects out the NSNS states. The operator that achieves this is
$\frac{1-(-1)^{F_L}}{2}$ as the NSNS sector states are invariant and
RR states change sign under $(-1)^{F_L}$. In what follows we carefully
analyze the effect of the insertion of $\frac{1-(-1)^{F_L}}{2}$ in the
tree channel on the loop channel amplitudes. We analyze the simple
case of the orientifold group $\{1, \Omega\}$ below. This can be
generalized to the other orientifold groups trivially. Let us consider
the Cylinder diagram first.

\subsubsection{The Cylinder}

Firstly recall that in the NSR formalism, the tree channel cylinder
amplitude can be thought of as the overlap of a boundary state $|Dp
\rangle$ representing a Dp-brane propagated for a distance $l$ with
itself and integrated over $l$,
\begin{equation}
\label{cbdry}
C~~~~:~~~~ \int_0^{\infty}dl\, \langle Dp| e^{-2\pi l\, H_c} |Dp
\rangle.
\end{equation}
This expression evaluates into the sum of contributions coming from
the exchange of NSNS sector fields and from the RR sector fields. If
we insert the operator ${ 1- (-1)^{F_L} \over 2}$ in (\ref{cbdry}), we
recover the contribution coming only from the exchange of the RR
sector fields alone. The fact that this is true in both NSR and GS
formalism is going to be exploited.
\bigskip
\FIGURE[h]{
{\epsfig{file=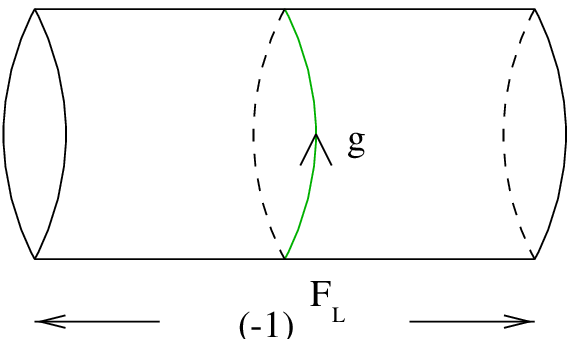, height=3.5cm}}
\caption{Cylinder diagram with $(-1)^{F_L}$ insertion in the tree
channel.}}
To find the RR contribution of the cylinder diagram we start with
the tree channel formula:
\begin{equation}
\label{cbdry2}
C~~~~:~~~~ \int_0^{\infty}dl\, \langle Dp|({1\over 2} - {(-1)^{F_L}
\over 2})\,e^{-2\pi\, l\, H_c} |Dp \rangle
\end{equation}
and translate it into a loop channel amplitude. We take the ranges of
the coordinates to be $0\leq \sigma^1\leq 2\pi l$, $0\leq \sigma^2\leq
2\pi $. The insertion of $1/2$ in (\ref{cbdry2}) gives half of the
amplitude in equation (\ref{cbdry}). On the other hand, the effect of
the insertion of $(-1)^{F_L}/2$ on the boundary conditions for the
fermions in the loop channel is as follows (see fig. (1)).  The
defining equation of the boundary state of the Dp-brane is,
schematically, given by:
\begin{equation}
\label{dpbcsone}
\left(S^a(\sigma^2) + i M_{ab} \tilde{S}^b(\sigma^2)\right)
|Dp \rangle = 0.
\end{equation}
This translated into the loop channel means that the open string
boundary condition{\footnote{Open string channel is obtained by
rescaling $\sigma^1$ and $\sigma^2$ by $2l$.}} is $S^a|_{\sigma^1} =
M_{ab}\tilde{S}^b|_{\sigma^1}$ where $M_{ab}$ is an operator made of
product of $SO(8)$ gamma matrices depending on the D-brane we want to
describe and $\sigma^1$ is either $0$ or $\pi$. Insertion of
$(-1)^{F_L}/2$ in the tree channel amounts to replacing the boundary
state $|Dp \rangle$ at $\sigma^1=\pi$ by $(-1)^{F_L}/2|Dp
\rangle$. Then Eq.(\ref{dpbcsone}) gives
\begin{eqnarray}
\label{dpbcstwo}
&&\left(S^a(\sigma^2) + i M_{ab}
\tilde{S}^b(\sigma^2)
\right) |Dp \rangle = 0\cr
&\Rightarrow& (-1)^{F_L}\left(S^a(\sigma^2) + i M_{ab}
\tilde{S}^b(\sigma^2)\right)(-1)^{2F_L}|Dp \rangle = 0\cr
&\Rightarrow& \left(S^a(\sigma^2) - i M_{ab}
\tilde{S}^b(\sigma^2)\right) (-1)^{F_L}|Dp \rangle = 0
\end{eqnarray}
where in the second line we have inserted $1=(-1)^{2F_L}$ and operated
by $(-1)^{F_L}$ from the left. Converting this into the open string
boundary conditions implies: $S^a|_{\sigma^1=\pi} = -
M_{ab}\tilde{S}^b|_{\sigma^1=\pi}$.  This makes the open string
fermions half integer moded.  That means the insertion of $(-1)^{F_L}$
in the tree channel picks the `wrong' boundary conditions in the loop
channel.{\footnote{One should note that the usual light-cone gauge
condition for the closed strings in GS formalism $X^+ \sim p^+ \tau$
allows for the boundary state description of only the instantonic
branes (transverse to the light-cone directions). On the other hand,
the same gauge choice for the open strings describes branes with
Neumann boundary conditions for the light-cone directions. To describe
the other branes, one has to choose a nonstandard light-cone gauge
$p^+ \sim x^+\tau$ \cite{bgg}. In our analysis here, we assume that we
are working in the nonstandard gauge in the closed string channel.}}

Therefore, we conclude that the RR contribution of the cylinder
diagram in the loop channel in GS formalism is
\begin{equation}
\label{onec}
\hbox{C}~~~:~~~ \hbox{\bf Tr}_{{\cal H}_o}
\left[\frac{1}{2} (-1)^{F_s} e^{-2\pi t\,\hbox{H}_o}\right]
- \hbox{\bf Tr}_{\tilde{\cal H}_o}
\left[\frac{1}{2} (-1)^{F_s} e^{-2\pi t\,\hbox{H}_o}\right],
\end{equation}
where we have denoted the open string Hilbert space of $8\times
\hbox{PB} \oplus 8 \times \hbox{AF}$ by $\tilde{\cal H}_o$ where PB
stands for a periodic boson and AF is an antiperiodic fermion.  The
operator $H_o$, which plays the role of $L_0$ (the Virasoro generator)
in NSR formalism, is given by:
\begin{equation}
\label{ho}
H_o = p^+ (p^- + H^o_{l.c})
\end{equation}
where $H^o_{l.c}$ is the light-cone Hamiltonian of the GS open
string. The trace $\hbox{\bf Tr}_{\tilde{\cal H}_o}$ in (\ref{onec})
includes an integration over the bosonic zero modes (the center of
mass coordinates and momenta) of the string and an integration
$\int_0^{\infty} dt/2t$ over the modular variable of the cylinder. The
operators inserted are simply determined by the requirement of the
usual periodic light-cone GS fermions.

Heuristically, the half-integer moding was to be expected. The
insertion of $(-1)^{F_L}$ converts $|Dp\rangle$ to
$|{\bar{Dp}}\rangle$ and the $D-\bar{D}$ overlap now will have a
tachyon in its spectrum which is consistent with half-integral moding.

\subsubsection{Klein Bottle and M\"obius Strip}

Let us now obtain the corresponding formulae for the RR contributions
to KB and MS diagrams. For this, it is useful to briefly review how to
translate the quantities from tree channel to loop channel in these
cases \cite{sagnotii,gimpol}.

Recall that to make a Klein bottle one can start with a torus with
coordinates $0\leq \sigma^1 \leq 4\pi l$ and $0\leq \sigma^2\leq 2\pi$
(with $\sigma^1 \sim \sigma^1 + 4\pi l$ and $\sigma^2 \sim \sigma^2 +
2\pi$) and make a $Z_2$ identification:
\begin{equation}
\label{z2}
(\sigma^1, \sigma^2) \sim ( 4\pi l - \sigma^1, \sigma^2 + \pi).
\end{equation}
One can choose two different fundamental regions and these two choices
naturally correspond to tree channel and loop channel string diagrams
(see figure 2).  The prescription to go from the tree to loop channel
is to (i) cut the upper half of tree channel diagram ($0\leq \sigma^1
\leq 2\pi l$ and $\pi \leq \sigma^2 \leq 2\pi$), (ii) invert it from
right to left, (iii) multiply the fields by $\Omega^{-1}$ ( by
$(\Omega h_2)^{-1}$ in the general case where the orientifolding
contains a space-time symmetry generator as well. See Fig.(3a)). and
(iv) glue it to the right side of the lower half.
\bigskip
\FIGURE[h]{
{\epsfig{file=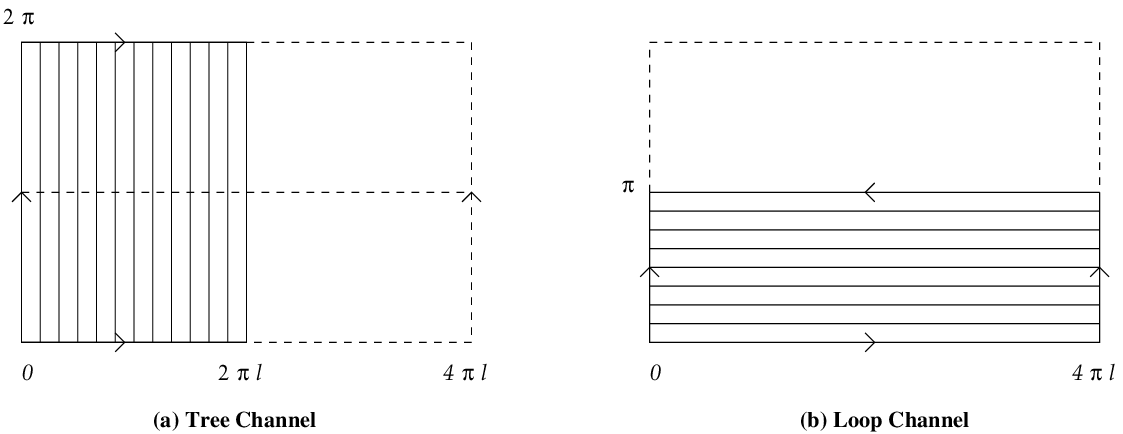, height=3.5cm}}
\caption{Two ways to view Klein Bottle}}
\bigskip
We are interested in going from the tree channel to loop channel when
there is a $(-1)^{F_L}$ operator inserted in the tree as one goes from
$0$ to $2\pi l$.  Since in this case, there is an operator
$(-1)^{F_L}$ in the upper half as well when we perform the step (iii)
above it will turn into $(-1)^{F_R}$ since the action of multiplying
by ${\cal O} \equiv \Omega^{-1}$ ($(\Omega {h_2})^{-1}$ in the general
case) on the fields becomes conjugation by ${\cal O}$ on the
operators. Therefore $(-1)^{F_L} \rightarrow {\cal O}^{-1}
(-1)^{F_L}{\cal O} = (-1)^{F_R}$. Hence by the end of step (iv) we
have the fields in the loop channel twisted by the operator
$(-1)^{F_L}\times (-1)^{F_R} = (-1)^{F_s}$. Since the loop channel of
KB is also a closed-string tree channel, both the left moving and
right moving GS closed string fermions $\tilde{S}^a$ and $S^a$ are
twisted by $(-1)^{F_s}$ giving rise to half-integer moding. Therefore,
the full RR contribution to the Klein Bottle diagram is
\begin{equation}
\label{onekb}
{\rm K.B}: ~~~~\hbox{\bf Tr}_{{\cal H}_c}
\left[\frac{\Omega}{2}(-1)^{F_s}\,e^{-2\pi t\,\hbox{H}_{c}}\right] 
- \hbox{\bf Tr}_{\tilde{{\cal H}}_c}
\left[\frac{\Omega}{2}(-1)^{F_s}\,e^{-2\pi t\,\hbox{H}_{c}}\right]
\end{equation}
where the first term comes from the insertion of $1/2$ and the second
one is from $(-1)^{F_L}/2$.
where $\tilde{{\cal H}}_c$ denotes the fock space of $(8\times
\hbox{PB} \oplus 8 \times \hbox{AF})\otimes (8\times \hbox{PB} \oplus
8 \times \hbox{AF})$. As in (\ref{ho}) $H_c = \frac{p^+}{2}(p^- +
2H^c_{l.c})$ with $H^c_{l.c}$ the closed string light-cone Hamiltonian.
\bigskip
\FIGURE[h]{
{\epsfig{file=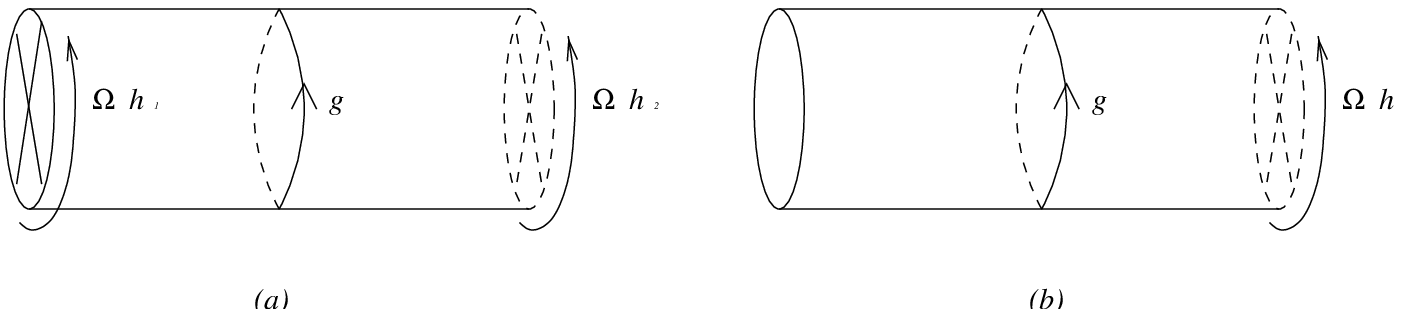, height=2.5cm}}
\caption{(a) Klein Bottle and (b) M\"obius Strip }}
\bigskip
Now we turn to the M\"obius Strip diagram. In this case one would have
started with a cylinder with coordinates $0\leq \sigma^1 \leq 4\pi l$
and $0\leq \sigma^2\leq 2\pi$ with $\sigma^2\sim\sigma^2+2\pi$ and
moded out by the same $Z_2$ action (\ref{z2}). The argument goes
through till we get an effective operator of $(-1)^{F_s}$ in the loop
channel. The difference in this case from that of KB is that in the
loop we have open strings. Therefore, using the method adopted for the
cylinder case, we look at the defining equation of the boundary state
$(-1)^{F_s}|Dp\rangle$ at $\sigma^1 = \pi$. It is easy to see from
this that the boundary conditions for the open string fermions at
$\sigma^1 = \pi$ are exactly the same as the ones without the
$(-1)^{F_L}$ insertion in the tree channel. Therefore, again we have
integer moded open string fermions.

Now to find the operator that is to be inserted in the open-string
trace it is convenient to consider the functions on the MS world-sheet
as functions of complex coordinates $z = \sigma^1 + i\, \sigma^2,
\bar{z} = \sigma^1 - i\, \sigma^2$. In terms of this the doubling
trick can be written as:
\begin{equation}
\label{dtagain}
\Psi (z + \pi) \equiv \Psi (\pi - \bar{z})
\end{equation}
and the action of $\Omega$ on the fields is given by
\begin{equation}
\label{omgactn}
\Omega \Psi(z) \Omega^{-1} = \tilde{\Psi}(\pi - \bar{z})
\end{equation}
Now suppose we have $R = \Omega \, e^{i\pi \beta F_s}$ in the
open-string trace. Then the field $\Psi$ gets twisted by $R$ across
the length of the strip in the $\sigma^2$ direction when translated by
$2\pi\,t$. Specifically on the fermions, we have (see (\cite{polch2})
for example)
\begin{eqnarray}
\label{msmanp}
S^a(z + i4\pi\,t) &=& -R\, S^a(z + i2\pi\,t)R^{-1} 
= - e^{i\pi\,\beta}S^a(\pi - \bar{z} + i2\pi\,t) \cr
&=& - e^{i\pi\,\beta}S^a(z + i2\pi\,t - \pi)
= S^a(z + 2\pi).
\end{eqnarray}
In going from the first line to the second we have made use of the
doubling trick (\ref{dtagain}). The equation (\ref{msmanp}) means that
we have periodic fermions in the tree channel irrespective of whether
$\beta = 0$ or $1$. Therefore we have to consider both the
possibilities. The one with $\beta = 1$ corresponds to the tree
channel amplitude with $1/2$ insertion and $\beta = 0$ to the one with
$(-1)^{F_L}/2$ insertion.  Therefore, for the RR contribution to the
MS diagram, we have:
\begin{equation}
\label{msone}
{\rm M.S}:~~~~\hbox{\bf Tr}_{{\cal H}_o}\left[\frac{\Omega}{2}\,
(-1)^{F_s} e^{-2\pi t \, \hbox{H}_o}\right]
-\hbox{\bf Tr}_{{\cal H}_o}\left[\frac{\Omega}{2}\,
e^{-2\pi t \, \hbox{H}_o}\right]
\end{equation}
where ${\cal H}_o$ is the Hilbert space of ($8\times {\rm PB} \oplus
8\times {\rm PF}$).
It is easy to check that these formulae (\ref{onec}), (\ref{onekb})
and (\ref{msone}) for the RR contributions of the diagrams C, KB and
MS calculated in GS formalism give rise to the answers
(\ref{rrtadone}) from NSR formalism for the orientifold type IIB on
${\mathbb R}^{9,1}$/$\{1, \Omega \}$ reviewed earlier in section
(2.1). All the above formulae are specific to the orientifold group
$\{1,\Omega\}$. These however can be generalized to other
orientifolding groups, ${\cal G}$ where
we replace $\Omega$ by $g_i\in{\cal G}$ and sum over $i$.

We conclude this section by pointing out an important difference
between the modular transformation in NSR formalism and the GS
formalism. Recall that in the GS formalism the light cone gauge is
fixed by choosing the light-cone direction $x^+ = \lambda p^+ \tau$
where $\lambda$ is a convention dependent factor ($\lambda = 2
\alpha'$ for us), $p^+$ is the light-cone momentum and $\tau$ is the
world-sheet time. Under a modular transformation $\tau \rightarrow
1/\tau$ we see that $p^+ \leftrightarrow x^+/\lambda$. Therefore using
$\tau = x^+/(\lambda p^+)$, we have $p^+ \rightarrow p^+ \tau$ as
observed in \cite{bgg}. So the set of modular transformations include
this change of $p^+$ along with the ones given in (\ref{modtrans}).

\section{Orientifolds of Plane-waves}

In this section we would like to orientifold the plane-wave geometry
given in (\ref{ppwave2}). The string world sheet propagating in this
background can be quantized in the light-cone gauge in the GS
formalism \cite{russotsey, yoji}. A review of the relevant aspects of
this quantization can be found in the appendix A.

\subsection{ Model I : Type IIB on $PP_6 \times {\mathbb
R}^4/\{1, \Omega\}$}

In the following, we evaluate the expressions (\ref{onec}),
(\ref{onekb}) and (\ref{msone}) for this orientifold. Firstly, we
define the action of the operator $\Omega$ on the various oscillator
modes as follows. For the closed string:
\begin{eqnarray}
\label{omegaoncs}
\Omega \, x^i(\sigma)\,\Omega^{-1} &=& x^i(2\pi - \sigma) 
~~~\,{\Rightarrow}~~~
\Omega~:~a_n^i \leftrightarrow {\tilde a}_n^i \\
\Omega \, s_L(\sigma)\,\Omega^{-1} &=& s_R(2\pi - \sigma) 
~~~{\Rightarrow}~~~
\Omega~:~s_m^a \leftrightarrow {\tilde s}_m^a 
\end{eqnarray}
%
For the open string
\begin{eqnarray}
\label{omegaonos}
\Omega \, x^i(\sigma)\,\Omega^{-1} &=& x^i(\pi - \sigma) 
~~~\,{\Rightarrow}~~~
\Omega~:~a_n^i \leftrightarrow (-1)^n a_n^i \\
\Omega \, s_L(\sigma)\,\Omega^{-1} &=& s_R(\pi - \sigma) 
~~~{\Rightarrow}~~~
\Omega~:~s_r^a \leftrightarrow (-1)^n s_r^a 
\end{eqnarray}
where the details of the oscillators can be found in the appendix
A. It is easy to check that the world-sheet action (\ref{saction})
written in these variables is invariant under $\Omega$. The action of
$\Omega$ on the ground states is $\Omega|0\rangle=|0\rangle$ for the
closed string vacuum of $\tilde{H_c}$ and
$\Omega|i/a\rangle=-|i/a\rangle$ for the open string vacua of
$H_o$. {\footnote{For the closed string the action is readily
understood since $\Omega$ interchanges right with left movers.  The
action of $\Omega$ on the ground states can be understood
heuristically as follows. For the $SO(8)$ vector ground state, the
corresponding vertex operator has a factor that depends on the tangent
derivative of $X^i$. Hence the action of $\Omega$ on this is to change
its sign. By virtue of $SO(8)$ triality, the spinor ground state is
related to the vector ground state by the identity
$S^a_0|i\rangle=\frac{1}{\sqrt
2}\gamma^i_{a\dot{a}}|\dot{a}\rangle$. Therefore, the action of
$\Omega$ on this is to reverse the sign of the spinor ground state as
well.}} This will lead to the M\"obius strip amplitude having a
negative sign with respect to the other two amplitudes.

\subsubsection{Tadpole Computation}

Let us now evaluate the loop amplitudes (\ref{onec}), (\ref{onekb})
and (\ref{msone}) in this model. The first terms of these expressions
vanish because of bose-fermi degeneracy at each level. The second
terms can be evaluated and expressed in terms of the modified modular
functions \cite{bgg} (see also \cite{takayanagi}) defined
as:{\footnote{These expressions are slightly different from those of
\cite{bgg}.}
\begin{eqnarray}
\label{newfs}
f_{1}^{(m)}(q)&=&q^{-2\Delta_m}(1-q^{2m})^{\frac{1}{2}}
\prod_{n=1}^{\infty}(1-q^{2\omega_n}), 
~~~f_{3}^{(m)}(q) = q^{-2\Delta'_m}\prod_{n=1}^{\infty}
(1+q^{2\omega_{n-\frac{1}{2}}}), \cr
f_{2}^{(m)}(q) &=& q^{-2\Delta_m}(1+q^{2m})^{\frac{1}{2}}
\prod_{n=1}^{\infty}(1+q^{2\omega_n}),
~~~f_{4}^{(m)}(q) = q^{-2\Delta'_m}\prod_{n=1}^{\infty}
(1-q^{2\omega_{n-\frac{1}{2}}})
\end{eqnarray}
where $q=e^{-2\pi t}$ and $\Delta_m$ and $\Delta^\prime_m$ are defined
by
\begin{eqnarray}
\label{deltas}
\Delta_m &=& -\frac{1}{(2\pi)^2} \sum_{p=1}^{\infty}
\int_0^\infty ds \, e^{-p^2 s} e^{-\pi^2 m^2 / s} \,, \cr
\Delta^\prime_m &=& -\frac{1}{(2\pi)^2} \sum_{p=1}^{\infty}
(-1)^p \int_0^\infty ds \, e^{-p^2 s} e^{-\pi^2 m^2 / s}.
\end{eqnarray}
See appendix B for a derivation of these from regularizing the
zero-point energy contributions.  Let us denote the regularised volume
of the six-dimensional non-compact space along $(+,-,5,6,7,8)$ by
$v_{6}$. After rescaling ${\hat p}^{-}=p^{-}t$ for open string and
${\hat p}^{-}=\frac{p^{-}}{2}t$ for the closed strings, the formulae
in the loop channel for the Cylinder, Klein-bottle and the M\"obius
strip are given by $v_{6}\int dp^{+} d{\hat p}^{-}e^{-p^+ {\hat p}^-}$
times {\footnote{One needs to regularise the $t$ integrals
$\int_0^\infty dt\,\rightarrow \int_\epsilon^\infty\, dt$ to make
sense of this change of variables. One can take $\epsilon \rightarrow
0$ after carrying out the integration over $p^+$ and ${\hat p}^-$.}}
\begin{equation}
\label{ansc}
\int_{0}^{\infty}\frac{dt}{2t^4}\left[f_{4}^{(m)}(e^{-\pi
t})f_{4}(e^{-\pi t})\over f_{1}^{(m)}(e^{-\pi t})f_1(e^{-\pi
t})\right]^4\frac{1}{(1-e^{-2\pi mt})^2}\times n^2
\end{equation}
for the cylinder. For the Klein bottle it is:
\begin{equation}
\label{anskb}
\int_{0}^{\infty}\frac{dt}{2t^4}2^3\left[f_{4}^{(m)}(e^{-2\pi
t})f_{4}(e^{-2\pi t})\over f_{1}^{(m)}(e^{-2\pi t})f_1(e^{-2\pi
t})\right]^4\frac{(1+e^{-2\pi t m})^2}{(1-e^{-2\pi mt})^2}.
\end{equation}
For the M\"obius strip it is:
\begin{equation}
\label{ansms}
- \int_{0}^{\infty}\frac{dt}{2t^4}\left[f_{4}^{(\frac{m}{2})}(e^{-2\pi
t})f_{4}(e^{-2\pi t})f_{2}^{(\frac{m}{2})}(e^{-2\pi t})f_2(e^{-2\pi
t})\over f_{1}^{(\frac{m}{2})}(e^{-2\pi t})f_1(e^{-2\pi
t})f_{3}^{(\frac{m}{2})}(e^{-2\pi t})f_3(e^{-2\pi
t})\right]^4\frac{1}{(1-e^{-2\pi mt})^2}\times n
\end{equation}
where $n$ denotes the number of D9 branes present and we have chosen
the $SO$ projection for the CP factors. Notice that the above
integrands resemble the obvious generalizations of the flat-space
results (\ref{rrtadone}) with 4 of the $f$'s replaced by $f^{(m)}$'s
along with the appropriate zero-mode contributions as indicated
above. One can now transform the integrals (\ref{ansc}, \ref{anskb},
\ref{ansms}) into tree-channel ones using the following modular
transformation properties of $f^{(m)}$'s and (\ref{fmodprop}) for
$f$'s.{\footnote{Notice that these modular transformations are
slightly different from the ones given in \cite{bgg}. One can show
that these are the correct relations for our definitions
(\ref{newfs}).}
\begin{eqnarray}
\label{newtrans}
f_{4}^{(m)}(e^{-2\pi t})&=&f_{2}^{(2mt)}(e^{-\frac{\pi}{2t}}), ~~~~~~~~
f_{2}^{(m)}(e^{-2\pi t}) = f_{4}^{(2mt)}(e^{-\frac{\pi}{2t}}), \cr
f_{1}^{(m)}(e^{-2\pi t})&=&f_{1}^{(2mt)}(e^{-\frac{\pi}{2t}}), ~~~~~~~~
f_{3}^{(m)}(e^{-2\pi t}) = f_{3}^{(2mt)}(e^{-\frac{\pi}{2t}}), 
\end{eqnarray}
Using these we get:
\begin{eqnarray}
\label{modoneans}
{\rm C}&:&~~~~~ \int_0^\infty \, dl ~ 
\left[\frac{f_2^{({\hat m})}(e^{-2\pi
l})f_2(e^{-2\pi l})}{f_1^{({\hat m})}(e^{-2\pi l})f_1(e^{-2\pi
l})}\right]^4  \frac{1}{(1- e^{-2\pi {\hat m}})^2}\times n^2 \cr
%
{\rm K.B}&:&~~~~\int_0^\infty \, dl ~ 
2^6 \left[\frac{f_2^{({\hat m})}(e^{-2\pi
l})f_2(e^{-2\pi l})}{f_1^{({\hat m})}(e^{-2\pi l})f_1(e^{-2\pi
l})}\right]^4  \left(\frac{1 + e^{-\pi {\hat m}}}
{1- e^{-\pi {\hat m}}}\right)^2 \\
%
{\rm M.S}&:&~~~~ - \int_0^\infty \, dl ~ 2^4 \left[\frac{f_2^{({\hat
m})}(e^{-4\pi l})f_2(e^{-4\pi l})f_4^{({\hat m})}(e^{-4\pi
l})f_4(e^{-4\pi l})}{f_1^{({\hat
m})}(e^{-4\pi l})f_1(e^{-4\pi l})f_3^{({\hat m})}(e^{-4\pi
l})f_3(e^{-4\pi l})} \right]^4 \frac{1}{(1- e^{-\pi{\hat m}})^2}\times n 
\nonumber
\end{eqnarray}
where the modular transformed mass parameters ${\hat m}$ are related
to loop channel ones via ${\hat m} = m/2l, \, m/2l,\, m/4l$ for
Cylinder, Klein-Bottle and M\"obius Strip respectively. 

The divergence in the RR contribution comes from the large $l$ limit
of the integrands. To cancel the RR tadpole we require that this
divergence to be absent. Here we still need to perform the
integrations $\int\, dp^+\, d{\hat p}^-\, e^{- p^+{\hat p}^-}$. There
is a subtlety in carrying this integral out since the loop channel
mass $m$ (and hence ${\hat m}$) for the world-sheet fields also
depends on $p^+$. As a result the integral over $p^{+},\hat{p}^{-}$
requires care. For this we make use of the following observation
\cite{GSW2}. It is possible to show that after Wick-rotating $p^0$,
\begin{equation}
\label{gsw2}
\int dp^{+}dp^{-}d^{8}\bar{p} (p^9+ip^0)^n
f(p^2)=\int_{0}^{2\pi}d\phi\,\int_{0}^{\infty}dr~ r^{n+1}e^{in\phi}\int
d^{8}\bar{p}f(\bar{p}^2+r^2), \label{formula}
\end{equation}
vanishes unless $n=0$ or the integral over $r$ is infinite. In the
right-hand-side of (\ref{gsw2}) we have substituted
$p^{9}+ip^{0}=re^{i\phi}$. In the integrals of our interest
(\ref{modoneans}) we typically encounter expressions of the form:
%
\begin{equation}
\label{typicalterm}
\int dp^{+}dp^{-}\frac{1}{(1-e^{-p^+ c})^2} e^{-p^+ p^-},
\end{equation}
where $c$ is independent of $p^+$ and $p^-$ and the integrand can be
formally expanded in powers of $p^+$. Now making use of the above
observation, we see that the only divergent contribution comes from
the leading order term proportional to $1/(p^{+})^2$ in the integrand
of (\ref{typicalterm}). That is, using equation (\ref{formula}), we
can transform the integral into the form,
\begin{equation}
\sum_{n=-2}^{\infty}\int_{0}^{2\pi}d\phi\, e^{in\phi}\int_{0}^{\infty}dr
r^{n+1}\, e^{-r^2},
\end{equation}
where the leading term is non-vanishing owing to the divergence of
$\int_{0}^{\infty}dr r^{-1}e^{-r^2}$.{\footnote{The terms
corresponding to $n = -1, 1,2,\cdots$ vanish trivially and the
contribution from $n=0$ integral is finite and subleading w.r.t the
$n=-2$ part.}} Therefore we see that the dominant contribution for our
integrals comes from the region very close to $p^+ =
0$.{\footnote{Another way of seeing this result is to take the limit
$l \rightarrow \infty$ first. This amounts to ${\hat m} \sim m/l
\rightarrow 0$ in the integrand which again leads to the conclusion
that the dominant contribution comes from the region ${\hat m}\sim
p^+/l \rightarrow 0$.}}

After using these relations the integrals in (\ref{ansc}, \ref{anskb},
\ref{ansms}) for large $l$ become $\int_0^\infty \, dl$ times
\begin{equation}
2^2 \frac{1}{(2\pi\hat{m})^2} \times n^2.
\end{equation}
For the Klein-bottle we get,
\begin{equation}
2^{12} \frac{1}{(2\pi\hat{m})^2}.
\end{equation}
Finally for the M\"obius Strip we get,
\begin{equation}
-2^8\frac{1}{(2\pi\hat{m})^2}\times n,
\end{equation}
Note that in the above integrals, $\hat{m}$ is the mass factor that
goes into the tree-channel amplitudes. Therefore, it is natural that
each of the integrals in the tree-channel need to be expressed in
terms of $\hat{m}$.{\footnote{We identify that the common factor of
$\left(\frac{1}{2\pi{\hat m}}\right)^2$ in the above integrals is
exactly the volume of the four dimensional space in
$(1,2,3,4)-$directions of the plane wave $PP_6$ in which string states
of light-cone momentum $p^+$ (from the tree channel point of view) get
confined.}
This leads to the result that for the vanishing of the RR tadpole,
\begin{equation}
(n-2^5)^2=0.
\end{equation}
This sets $n=32$ corresponding to the gauge group $SO(32)$ for the
corresponding type I superstring in this background.

\subsubsection{The Spectrum}
Having seen that we need 32 D9-branes for tadpole consistency, we now
would like to spell out the low energy spectrum of the resultant
theory. Since the isometry group of (\ref{ppwave2}) is $SO(2)\times
SO(2)\times SO(4)$ the physical states of the strings should belong to
the representations of this group. In fact it is possible to identify
various supergravity modes in this background with the corresponding
low energy string states following \cite{mettsey,atishparv}. In the
following we present this analysis of light-modes coming from the
open-strings alone.

The supergravity modes correspond to the string states obtained by
acting with the zeromodes $a_0^i$ and $s_0^a$ (and their massless
counterparts coming from the four flat directions transverse to the
plane wave) on the vacuum state defined in the appendix A.
We use the following embedding \cite{atishparv}:
\begin{eqnarray}
\label{embed}
SO(8) &\supset & \widetilde{SU}(2)_L\times \widetilde{SU}(2)_R \times
SU(2)_1\times SU(2)_2\cr
&\supset & SO(2)_1\times SO(2)_2\times SU(2)_1\times SU(2)_2.
\end{eqnarray}
Notice that $SO(2)_1\times SO(2)_2$ is the group of rotations in 12
and 23 planes generated by $T_{12}$ and $T_{34}$ respectively. In
terms of the generators of $\widetilde{SU}(2)_L\times
\widetilde{SU}(2)_R$ these are given by:
\begin{eqnarray}
\label{rotgendefs}
T_{12} &=& \tilde{J}_{3L} + \tilde{J}_{3R}\cr
T_{34} &=& \tilde{J}_{3L} - \tilde{J}_{3R}
\end{eqnarray}
Under this embedding the spinor decomposes as:
\begin{equation}
\label{decomp}
{\bf 8}_{\rm s} \sim ({\bf 2}, {\bf 1})^{(\frac{1}{2}, 
\frac{1}{2})} \oplus 
({\bf \bar{2}}, 1)^{(-\frac{1}{2}, -\frac{1}{2})} \oplus 
({\bf 1}, {\bf 2})^{(\frac{1}{2}, - \frac{1}{2})} \oplus
({\bf 1}, {\bf \bar{2}})^{(-\frac{1}{2}, \frac{1}{2})}
\end{equation}
where the superscript denotes the $SO(2)_1\times SO(2)_2$
charges. Using this decomposition, we can now organize the zero modes
of the spinor in terms of fermionic creation and annihilation
operators:
\begin{eqnarray}
\label{crandefs}
\bar{\lambda}_\alpha = S_{0\alpha}^{(\frac{1}{2}, \frac{1}{2})}, &~~~&
\lambda_\alpha = S_{0\alpha}^{(-\frac{1}{2}, -\frac{1}{2})}, \cr
\bar{\lambda}_{\dot{\alpha}} = S_{0\dot{\alpha}}^{(-\frac{1}{2}, 
\frac{1}{2})}, &~~~&
{\lambda}_{\dot{\alpha}} = S_{0\dot{\alpha}}^{(\frac{1}{2}, 
-\frac{1}{2})}.
\end{eqnarray}
where $\alpha$ and $\dot{\alpha}$ are the doublet indices of $SU(2)_1$
and $SU(2)_2$ respectively. Now we want to write the Hamiltonian in
terms of these modes in order to be able to build the spectrum. Let us
start with the open string Hamiltonian. As given in the appendix the
zero-mode piece in the Hamiltonian (\ref{openlcham}) relevant for the
D9-brane case is
\begin{equation}
E_{0}=\frac{1}{p^+}({\rm bosons} -\frac{im}{2}{\alpha'} s_0 \Lambda
s_0 + 2\omega_0/\alpha')
\end{equation}
where we recall that $s_0$ is the zero-mode of
$\frac{1}{2}(1-\gamma^{1234})S$, $S$ being and $SO(8)$ spinor and
$\Lambda$ is the upper diagonal component of $\gamma^{12}$. The
$2\omega_0$ comes from the fact that there are 4 bosonic harmonic
oscillators each contributing $\omega_0/2 = m/2$ to the normal
ordering. Rewriting the above equation in terms of $SO(8)$ variables,
we get
\begin{equation}
E_0= 2f({\rm bosons} + 2 - \frac{i}{2}S_0(\gamma^{12}+\gamma^{34})S_0),
\end{equation}
where $\gamma^{12}$ and $\gamma^{34}$ are proportional to the rotation
generators in the $1,2$ and $3,4$ planes respectively. As a result the
zero-mode part of the Hamiltonian becomes in terms of $\lambda$'s,
\begin{equation}
E_0 = 2f({\rm bosons} - 2\bar{\lambda}_{\alpha}\lambda^{\alpha}+4),
\end{equation}
where we have used
\begin{equation}
\label{algebra}
\{\bar{\lambda}_{\alpha},\lambda^{\beta}\}=\delta_{\alpha}^{\beta}.
\end{equation}
Notice that that the $SU(2)_R$ spinors do not appear in the formula
above. This is to be expected since in our case, we have 4 fermionic
zero modes which can be used to make a 2-component raising and a
2-component lowering operator which we identify to be
$\bar{\lambda}_{\alpha}$ and ${\lambda}_{\alpha}$ respectively. Let us
define the Fock vacuum for the Clifford algebra (\ref{algebra}) to be
the state with $T_{12}$ charge $0$ and $T_{34}$ charge $-1$ and
annihilated by $a_0$ the bosonic zero mode annihilation operator,
$\lambda^\alpha$ and $\lambda^{\dot{\alpha}}$. We are now ready to
describe the open string zero mode spectrum which we summarize in
table (1) below.
\TABLE[h]{
\begin{tabular}{rlcc}
State & Representation & ${{\rm Energy} \atop E_0/2f}$ & Field \\
$|0, -1~\rangle$ & $({\bf 1},{\bf 1})^{(\,~0,-1)}$ & 4 & $\bar{B}$\\
${\bar \lambda}_\alpha|0, -1~\rangle$ & $({\bf 2},{\bf
1})^{(~\frac{1}{2},-\frac{1}{2})}$  
& 2 & $\psi_{\alpha}$\\
$\bar{\lambda}_{\dot{\alpha}}|0,-1~\rangle$ 
& $({\bf 1},{\bf 2})^{(-\frac{1}{2},-\frac{1}{2})}$ 
& 4 & $\psi_{\dot{\alpha}}$ \\
${\bar \lambda}_\alpha{\bar \lambda}_\beta |0,-1~\rangle$ 
& $({\bf 1},{\bf 1})^{(\,~1, \,~0)}$ & 0 & $A$\\
${\bar \lambda}_\alpha\bar{\lambda}_{\dot\alpha}|0,-1~\rangle$
& $({\bf 2},{\bf 2})^{(\,~0, \,~0)}$ & 2 & $A_i$\\
${\bar \lambda}_{\dot{\alpha}}{\bar \lambda}_{\dot\beta} |0,-1~\rangle$ 
& $({\bf 1},{\bf 1})^{(-1, ~0)}$ & 4 & $\bar{A}$ \\
${\bar \lambda}_\alpha\bar{\lambda}_{\dot{\alpha}}{\bar
\lambda}_{\dot\beta}|0,-1~\rangle$ 
& $({\bf 2},{\bf 1})^{(-\frac{1}{2}, ~\frac{1}{2})}$ & 2 
& $\bar{\psi_{\alpha}}$ \\
${\bar \lambda}_\alpha{\bar \lambda}_\beta\bar{\lambda}_{\dot{\alpha}}
|0,-1~\rangle$ 
& $({\bf 1},{\bf 2})^{(~\frac{1}{2}, ~\frac{1}{2})}$ & 0 
& $\bar{\psi}_{\dot{\alpha}}$ \\ 
${\bar \lambda}_\alpha{\bar \lambda}_\beta \bar{\lambda}_{\dot{\alpha}}
{\bar \lambda}_{\dot\beta} |0,-1~\rangle$ 
& $({\bf 1},{\bf 1})^{(\,~0, \,~1)}$ & 0 & B
\end{tabular}
\caption{Light open string states and their corresponding world-volume
fields.}} 
Let us now identify the gauge field fluctuations in the table above.
The low energy effective theory on the 32 D9-branes contains an SO(32)
gauge field $A_M$ and the gauginos. The dynamics of this gauge field
is governed by the action:
\begin{equation}
\label{dnineaction}
S^{(D9)} = {\rm Tr}\, \int_{PP_6\times R^4} \frac{1}{2}F\wedge *F +
\frac{1}{4!} C^{(2)}\wedge F\wedge F\wedge F\wedge F + \frac{1}{2!}
C^{(6)}\wedge F\wedge F
\end{equation}
where $C^{(2)}$ is the RR 2-form and the $C^{(6)}$ is its magnetic
dual (i.e, $F^{(3)} = dC^{(2)} = *dC^{(6)} = *F^{(7)}$). The equation
of motion that we get from this action for the quadratic fluctuations
is:
\begin{equation}
\label{eomfora}
d*F = F^{(7)}\wedge F.
\end{equation}
Working in the light cone gauge the transverse gauge field $A_M$
decomposes, under the isometry group (\ref{embed}), into:
\begin{equation}
A_i\,({\bf 2}, {\bf 2})^{(0,0)}\oplus A\,({\bf 1}, {\bf 1})^{(1,0)}
\oplus {\bar A}\,({\bf 1}, {\bf 1})^{(-1,0)}
\oplus B\,({\bf 1}, {\bf 1})^{(0,1)}
\oplus {\bar B}\,({\bf 1}, {\bf 1})^{(0,-1)}
\end{equation}
In terms of these fields the equation of motion can be rewritten
as:
\begin{equation}
\left[ \Box - 2i(2f)(T^{12} + T^{34})\partial_- \right] A_r = 0
\end{equation}
where $A_r$ are the transverse components of the $SO(32)$ gauge
fields. The normal ordered Hamiltonian becomes:
\begin{equation}
\label{gfham}
H = 2f\left[N_b + 2 - 2(T^{12} + T^{34})\right].
\end{equation}
This explains the bosonic field content in the last column of the
table (1) above. Note that we have Bose-Fermi degeneracy at each
level. This is expected as we have the zero mode oscillators coming
from the flat directions in our plane-wave geometry which do not raise
the light-cone energy. This fact also
reflects in the vanishing of the supersymmetric one loop amplitudes.

\subsection{Model II : Type IIB on $PP_6 \times T^4/\{1,
\Omega I_{5678}\}$}

We now consider type IIB on $T^{4}/\{1,\Omega I_{5678}\}$. This model
is T-dual to model I. The action of
$\Omega I_{5678}$ on the open-string oscillator modes is as
follows. For bosonic oscillators,
\begin{eqnarray}
a_{-n}^{i}&\rightarrow& (-1)^{n+1} a^{i}_{-n}~~~ {\rm if}~ i = 5, 6, 
7, 8\\
&\rightarrow& (-1)^n a^{i}_{-n}~~~~~~ {\rm otherwise}
\end{eqnarray}
For fermionic oscillators,
\begin{equation}
S^{a}_{-n}\rightarrow (-1)^n P_5 P_6P_7 P_8 S^{a}_{-n}
\end{equation}
where $P_{i}=\gamma \gamma^{i}$. The action of $\Omega I_{5678}$ on
the closed-string oscillator modes is similar to the ones above
without the $(-1)^n$ and with
left and right moving oscillators interchanged. $\Omega$ and
$R=I_{5678}$ have the following action on the momenta and winding
modes.
\begin{equation}
R|p,w>=|-p,-w>\quad \Omega|p,w>=|p,-w>
\end{equation}
where $p,w$ label the compact momenta and winding respectively.

The tadpole in this orientifold can be extracted in a similar manner
as in the previous case. In what follows, the integration over $p^+,
p^-$ will be implicit. The contributions to the tadpole are given by
$v_2\int \frac{dt}{2 t^2}$ times

\begin{eqnarray}
{\rm C:} &&\left(\frac{f_{4}^{(m)}(e^{-\pi t})f_{4}(e^{-\pi
t})}{f_{1}^{(m)}(e^{-\pi t})f_{1}(e^{-\pi t})}\right)^4 \sum_{i,j\in
5}(\gamma_{1,5})_{ii}(\gamma_{1,5})_{jj} \\ \nonumber 
&& \times \prod_{m=5}^{8}\sum_{w=-\infty}^{\infty}
e^{-\frac{t}{2\pi\alpha'}(2\pi w r+x_{i}^{m}-x_{j}^{m})^2}
\frac{1}{(1-e^{-2\pi m t})^2}\\
{\rm KB:}&&2 \left(\frac{f_{4}^{(m)}(e^{-2\pi t})f_{4}(e^{-2\pi
t})}{f_{1}^{(m)}(e^{-2\pi t})f_{1}(e^{-2\pi
t})}\right)^4\left(\sum_{w=-\infty}^{\infty}e^{-\pi t \rho
w^2}\right)^4\left(\frac{1+e^{-2\pi m t}}{1-e^{-2\pi m t}}\right)^2 \\
{\rm MS:}&& -2^2 \left(\frac{f_{4}^{(\frac{m}{2})}(e^{-2\pi t})
f_{2}^{(\frac{m}{2})}(e^{-2\pi
t})}{f_{1}^{(\frac{m}{2})}(e^{-2\pi t})f_{3}^{(\frac{m}{2})}(e^{-2\pi
t})}\right)^4\frac{1}{\left(f_{1}(e^{-2\pi t})f_{3}(e^{-2\pi
t})\right)^4 }\\ \nonumber 
&&\times\left(\sum_{w=-\infty}^{\infty}
e^{-2\pi t \rho w^2}\right)^4{\rm Tr}(\gamma^{-1}_{\Omega
R,5}\gamma^{T}_{\Omega R,5})\frac{1}{(1-e^{-2\pi m t})^2}
\end{eqnarray}

The reason for the appearance of the zero mode factors in the various
diagrams should be evident from the analysis of the previous model. In
the above formulae, $\gamma_g$ is a matrix associated with the element
$g$ in the orientifolding group.  Transforming the above integrals
into tree-channel and performing a Poisson resummation for the winding
modes, using the formula,
\begin{equation}
\sum_{n=-\infty}^{\infty}e^{-\pi
(n-b)^2/a}=\sqrt{a}\sum_{s=-\infty}^{\infty}e^{-\pi as^2+2\pi isb},
\end{equation}
we get $\frac{v_2}{v_4}\int
dl$ times
\begin{eqnarray}
{\rm C:} &&2^{-2}\left(\frac{f_{2}^{(\hat{m})}(e^{-2\pi l})
f_{2}(e^{-2\pi l})}{f_{1}^{(\hat{m})}(e^{-2\pi l})f_{1}
(e^{-2\pi l})}\right)^4 \sum_{i,j\in 5}(\gamma_{1,5})_{ii}
(\gamma_{1,5})_{jj}\\ \nonumber
&&\times\prod_{m=6}^{9}\sum_{s=-\infty}^{\infty}
e^{-\frac{-\pi l \alpha' s^2}{r^2}
+ is\frac{(x_{i}^{m}-x_{j}^{m})}{r}}
\frac{1}{(1-e^{-2\pi\hat{m}})^2}\\
{\rm KB:}&&2^4 \left(\frac{f_{2}^{(\hat{m})}
(e^{-2\pi l})f_{2}(e^{-2\pi l})}{f_{1}^{(\hat{m})}
(e^{-2\pi l})f_{1}(e^{-2\pi l})}\right)^4
\left(\sum_{s=-\infty}^{\infty}
e^{-\frac{4\pi s^2 l}{\rho}}\right)^4
\left(\frac{1+e^{-\pi\hat{m}}}{1-e^{-\pi\hat{m}}}\right)^2 \\
{\rm MS:}&& -2^4 \left(\frac{f_{4}^{(\hat{m})}
(e^{-4\pi l})f_{2}^{(\hat{m})}(e^{-4\pi l})}{f_{1}^{(\hat{m})}
(e^{-4\pi l})f_{3}^{(\hat{m})}(e^{-4\pi l})}\right)^4
\frac{1}{\left(f_{1}(e^{-4\pi l})f_{3}(e^{-4\pi l})\right)^4 }
\\ \nonumber
&&\times\left(\sum_{s=-\infty}^{\infty}
e^{-\frac{4\pi s^2 l}{\rho}}\right)^4{\rm Tr}
(\gamma^{-1}_{\Omega R,5}\gamma^{T}_{\Omega R,5})
\frac{1}{(1-e^{-\pi\hat{m} })^2}
\end{eqnarray}
where $\rho=r^2/\alpha'$, $v_4=\alpha'^{2}/r^{4}$ and the $\hat{m}$'s
are as in the previous example.

Taking the asymptotics of the above leads to
$\frac{v_2}{v_4\pi^2\hat{m}^2}$
\begin{equation}
{\rm C~:}~~n^2 2^{-2}, ~~~ {\rm KB~:}~~ 2^8, ~~~
{\rm MS~:}~~ -2^4 n
\end{equation}
where $n={\rm Tr}(\gamma_{1,5})={\rm Tr}(\gamma^{-1}_{\Omega
R,5}\gamma^{T}_{\Omega R,5})$ is the number of D5 branes needed to
cancel the tadpole. This leads to the condition that 32 D5
branes are required. 

\subsection{The Heterotic Duals}

We have seen that there are consistent orientifolds of the background
(\ref{ppwave2}). The first model of Type IIB on $PP_6 \times {\mathbb
R}^4$/ $\{1, \Omega \}$ is now a background of the type I theory with
SO(32) gauge group:
\begin{eqnarray}
\label{ppwave3}
ds^2 &=& dx^+dx^- - f^2x^ix^i (dx^+)^2 
+ dx^idx^i + dx^\alpha dx^\alpha \cr 
&& ~~~ F_{+12} = F_{+34} = -2f, ~~~ \Phi = A_\mu = 0.
\end{eqnarray}
Similarly the second model of type IIB on $PP_6 \times T^4$/ $\{1,
\Omega I_{5678} \}$ is also a solution of the type I supergravity with
appropriate Wilson lines in SO(32) gauge group turned on. These Wilson
lines depend on the positions of the D5 branes on $T^4$. 

As a background of type IIB, the solution (\ref{ppwave2}) is a
supersymmetric one. Since it can be obtained as a Penrose limit of
$AdS_3\times S^3\times T^4$, it should admit a minimum of 16
supersymmetries. But in fact this background (\ref{ppwave2}) turns out
to have 24 supersymmetries. The 8 additional supersymmetries are the
so called `supernumerary' ones. Ignoring these for the time being, our
backgrounds break half of the dynamical suspersymmetries and half of
the kinematical supersymmetries.

It is natural to look for the possible heterotic duals of the
backgrounds of type I theory that we have found by orientifolding the
type IIB theory.

Let us first consider the S-dual of our first model. Recall that the
S-duality involves the following field redefinitions:
\begin{equation}
\Phi^{(I)} = - \Phi^{(H)}, ~~~ g_{\mu\nu}^{(I)} 
= e^{- \Phi^{(H)}}g_{\mu\nu}^{(H)}, ~~~ B_{\mu\nu}^{(I)} =
B_{\mu\nu}^{(H)}~~ {\rm and} ~~ A_\mu^{(I)} = A_\mu^{(H)}.
\end{equation}
Since the dilaton $\Phi^{(I)}$ is trivial in our type I background
we have exactly the same supergravity solution for the heterotic dual
with the B-field of type I replaced by that in the heterotic. The fact
that the number of supersymmetries and the low energy spectra
match is also not surprising. We hope to return to the question of
more detailed checks in future.

\section{Conclusions}

We have set up a formalism to compute RR tadpoles using the
Green-Schwarz string in the light-cone gauge. Using it, we studied the
tadpoles of two orientifold models of the plane wave obtained as the
Penrose limit of $AdS_3\times S^3\times T^4$ with a null RR three form
flux. We have seen that the RR tadpoles require 32 D-branes in the
background similar to the Minkowski background. This result can be
understood as follows. One could have started with $N_1$ D1-branes and
$N_5$ D5-branes parallel to each other and taken the near-horizon
limit and then Penrose limit. This would have given us the background
of the result of our first model. Similarly the second model can also
be obtained this way. It is satisfying to see that we are able to
recover these backgrounds as orientifolds of the corresponding type
IIB backgrounds. We have studied the spectrum of light open strings in
detail. We have also proposed the possible heterotic duals for our
models.

It would be interesting to see if our method can be used to construct
other orientifolds of plane waves. For example, one could start with
the maximally supersymmetric plane-wave of type IIB, compactify on a
$T^2$ on the lines of \cite{jm} and orientifold it using the group
$\{1, \Omega (-1)^{F_L} I_2 \}$, where $I_2$ is the reflection along
the $T^2$ directions. The study of this model is currently in
progress. Further, how these models fit into the string-string duality
network and how the string-CFT dualities work is also an open problem.

It would also be interesting to see if these backgrounds admit nice
holographic descriptions \cite{sugawara,gmstrom,gavanara} in terms of
the dual 2+1 dimensional CFT's.{\footnote{ See, for example,
\cite{ksn1} for holographic description of the near horizon geometries
of D1-D5 systems in the orientifold models.}}  We hope to return to
these issues in the future.

\bigskip
\noindent{\bf Acknowledgements}: 
We thank Pascal Bain, Justin David, Marta G\'omez-Reino, Stefano
Kovacs, Carlos N\'u\~nez, and especially Atish Dabholkar, Michael
Green and Ashoke Sen for valuable discussions. Special thanks to
Pascal Bain and Michael Green for going through the draft. AS is
supported by the Gates Cambridge scholarship and the Perse scholarship
of Gonville and Caius college. NVS is supported by PPARC Research
Assistantship.

\appendix
\small
\baselineskip 12pt
\section{Review of String Quantization in the plane-wave Background}
The action for the strings in the background of (\ref{ppwave2}) is
given by{\footnote{we use a convention where the closed string world
sheet has $0\le \sigma \le 2\pi$ and for the open string $0\le \sigma
\le \pi$.}}:
\begin{eqnarray}
\label{saction}
S =&& {1\over \pi \alpha'}\int d^2\sigma\, \Bigg[{1\over 2}\partial_+x^+
\partial_-x^-+ {1\over 2}\partial_+x^- \partial_-x^+ 
-{ m^2\over 4} x^i x^i + \partial_+ x^i \partial_-x^i + \partial_+ 
x^a \partial_- x^a \cr
&&~~~~~~~~~~~~~~~~+ iS_R \partial_+ S_R + i S_L \partial_- S_L 
+ imS_L \gamma^1\gamma^2 R_+ S_R \Bigg]
\end{eqnarray}
where $\gamma^i$ are the SO(8) real gamma matrices and $R_{\pm} =
\frac{1}{2}(1\mp \gamma^1\gamma^2\gamma^3\gamma^4)$. Choosing the
light cone gauge $x^+ = 2\alpha' p^+ \tau$ and $m = 2\alpha' p^+ f =
4\alpha' p_- f$. Following \cite{russotsey, yoji} we decompose
$S_{R,L}$ into four component spinors $s_{R,L}$ and $\hat{s}_{R,L}$:
\begin{equation}
S_{R,L} = \left({s_{R,L}\atop {\hat s}_{R,L}}\right), ~~~R_+S_{R,L} =
\left({s_{R,L}\atop 0}\right),~~\gamma^{12}S_{R,L} = - \left({\Lambda
s_{R,L}\atop {\hat \Lambda}{\hat s}_{R,L}}\right)
\end{equation}
where $\Lambda$ and ${\hat \Lambda}$ are $4\times 4$ antisymmetric
matrices with $\Lambda^2 = {\hat \Lambda}^2 = -1$.

\bigskip
\noindent{\underbar{\bf Closed Strings}}:

The equations of motion for $x^i$ and the fermions following from
(\ref{saction}) are
\begin{eqnarray}
\label{eoms}
\partial_+ \partial_- x^i + {m^2 \over 4}x^i &=& 0, 
~~~\partial_+ \partial_- x^\alpha = 0, ~~~\partial_+ s_R - {m \over
2}\Lambda s_L = 0, \cr 
\partial_- s_L - {m \over 2}\Lambda s_R &=& 0,
~~~~\partial_+ \hat{s}_R = 0,
~~~~~~\partial_- \hat{s}_L = 0.
\end{eqnarray}
The general solutions to these equations of motion are:
\begin{eqnarray}
\label{csolnsx}
x^i(\sigma, \tau) &=& i \sqrt{{\alpha'\over 2}} [ {1\over
\sqrt{\omega_0}}(a_0^i e^{-i\omega_o\tau} - a_0^i{}^\dagger
e^{i\omega_o\tau}) \cr
&& + \sum_{n=1}^{\infty} {1\over
\sqrt{\omega_n}}[e^{-i\omega_n \tau} ( a_n^i e^{in\sigma} +
\tilde{a}_n^i e^{-in\sigma}) - e^{i\omega_n \tau}(a_n^i{}^\dagger
e^{-in\sigma} + \tilde{a}_n^i{}^\dagger e^{in\sigma}) ]
\end{eqnarray}
where
\begin{equation}
\label{omgdef}
\omega_n = \sqrt{n^2 + m^2}, ~~~~ \omega_0 = m.
\end{equation}
The canonical quantization leads to the following commutation relation
for the modes:
\begin{equation}
\label{ccomx}
[a_n^i, a_m^j{}^\dagger ] = \delta_{m,n}\delta^{i,j}, ~~~~
[\tilde{a}_n^i, \tilde{a}_m^j{}^\dagger ] = \delta_{m,n}\delta^{i,j}
\end{equation}
For the fermions the solutions are
\begin{eqnarray}
\label{cfsoln}
s^a_R(\sigma, \tau) &=& \sqrt{2\alpha'}\Big[c_0(s_0^a
e^{-i\omega_0 \tau} - s_0^a{}^\dagger e^{i\omega_0 \tau}) + \sum
_{n=1}^{\infty} c_n [ e^{-i\omega_n\tau}(s_n^ae^{in\sigma} +
\tilde{s}'^a_n e^{-in\sigma}) \cr
&& ~~~~~~ + e^{i\omega_n\tau}(s_n^a{}^\dagger e^{-in\sigma} +
\tilde{s}'^a_n e^{in\sigma})] \Big] \cr
s_L^a(\sigma, \tau) &=& \sqrt{2\alpha'}\Big[ (\tilde{s}_0^a
e^{-i\omega_0 \tau} + \tilde{s}_0^a{}^\dagger e^{i\omega_0 \tau}) + \sum
_{n=1}^{\infty} c_n [ e^{-i\omega_n\tau}(\tilde{s}_n^a e^{in\sigma} +
s'^a_n e^{-in\sigma}) \cr
&& ~~~~~~ + e^{i\omega_n\tau}(\tilde{s}_n^a{}^\dagger e^{-in\sigma} +
s'^a_n e^{in\sigma})] \Big]
\end{eqnarray}
The primed fermionic modes are related to the unprimed ones
\cite{russotsey}. The canonical quantization gives the following
commutation relations.
\begin{equation}
\label{cfcomm}
\{s_m^a, s_n^b{}^\dagger \} = \delta^{a,b}\delta_{m,n}, ~~~
\{\tilde{s}_m^a, \tilde{s}_n^b{}^\dagger \} = \delta^{a,b}\delta_{m,n},
~~~ \{s_0^a, s_0^b{}^\dagger \} = \delta^{a,b}
\end{equation}
where we have fixed the coefficients $c_0$ and $c_n$ to be
\begin{equation}
c_0 = {1\over \sqrt{2}}, ~~~~~~~~ c_n = {m \over \sqrt{m^2 + (\omega_n
- n)^2}}.
\end{equation}
The vacuum $|0\rangle$ can be defined to be annihilated by the
positive frequency modes $a_n^i, \tilde{a}^i_n$ and $s_n^i,
\tilde{s}_n^a$ with $n = 1,2,..., \infty$ as well as the zero modes
$a_0^i$ and $s_0^a$. Along with these four massive bosons and the four
massive fermions we have four massless bosons $x^\alpha$ and four
massless fermions as well which we do not write them here as they are
standard.

The light cone Hamiltonian is given by 
\begin{equation}
\label{cham}
H_{l.c} = - p_+ = {p_\alpha^2 \over 4p_-} 
+ {1\over 2\alpha' p_-}{\cal H},~~ {\cal H} = {\cal H}_0 
+ {\cal H}_R + {\cal H}_L + N_R^0 + N_L^0,
\end{equation}
where 
\begin{equation}
{\cal H}_0 = \omega_0(a_0^i{}^\dagger a_0^i + s_0^a{}^\dagger s_0^a),
\end{equation}
\begin{equation}
{\cal H}_R = \sum_{n=1}^{\infty} \omega_n(a_n^i{}^\dagger a_n^i +
s_n^a{}^\dagger s_n^a), ~~~~~ 
{\cal H}_L = \sum_{n=1}^{\infty} \omega_n(\tilde{a}_n^i{}^\dagger 
\tilde{a}_n^i + \tilde{s}_n^a{}^\dagger \tilde{s}_n^a),
\end{equation}
\begin{equation}
N_R = \sum_{n=1}^{\infty} n(a_n^i{}^\dagger a_n^i +
s_n^a{}^\dagger s_n^a), ~~~~~ 
N_L = \sum_{n=1}^{\infty} n(\tilde{a}_n^i{}^\dagger 
\tilde{a}_n^i + \tilde{s}_n^a{}^\dagger \tilde{s}_n^a),
\end{equation}
\begin{equation}
N_R^0 = \sum_{n=1}^{\infty} n(a_n^\alpha{}^\dagger a_n^\alpha +
s_n^A{}^\dagger s_n^A), ~~~~~ 
N_L^0 = \sum_{n=1}^{\infty} n(\tilde{a}_n^\alpha{}^\dagger 
\tilde{a}_n^\alpha + \tilde{s}_n^A{}^\dagger \tilde{s}_n^A).
\end{equation}
The physical state condition is $N_R + N_R^0 = N_L + N_L^0$. The
string states are then constructed by acting by the creation operators
on the vacuum. The theory admits $SO(2)\times SO(2)\times SO(4)$
symmetry which is the isometry group of the solution
(\ref{ppwave2}). The states of the Hilbert space fall into
representations of this group.

\bigskip
\noindent{\underbar{\bf Open String}}:

The boundary conditions for the open string at $\sigma=0,\pi$ are
\begin{equation}
{\rm N}~:~\partial_{\sigma}x^{i}=0,~~~~ {\rm D}~:~x^{i}={\rm const.},~~~
S_{L} =  MS_{R}~~{\rm and}~~ M^{T}M=1.
\end{equation}
In the light-cone gauge $x^{+}$ and $x^{-}$ are always Neumann. The
possible boundary conditions defining various D-branes are studies in
detail in \cite{yoji}. In this case of both ends being Neumann, the
mode expansion for $x^{i}$ is given by
\begin{equation}
x^i=i\sqrt{\frac{\alpha'}{2}}{\Bigg [}\sqrt{2\over\omega_{0}}
(a_{0}^{i}e^{-i\omega_{0}\tau}
-a_{0}^{i\dagger}e^{i\omega_0\tau}) +
\sum_{n=1}^{\infty}\frac{1}{\sqrt{\omega_{n}}}
(a_{n}^{i}e^{-i\omega_{n}\tau}-a_{n}^{i\dagger}e^{i\omega_n\tau})
(e^{in\sigma}+e^{-in\sigma}){\Bigg ]},
\end{equation}
where 
\begin{equation}
\omega_{n}={\rm sgn}(n)\sqrt{n^2+m^2}, \quad (n\neq 0), \quad
\omega_0=m, 
\end{equation}
\begin{equation}
[a^{i}_{n},a^{j\dagger}_{n'}]=\delta^{i,j}\delta_{n,n'},
\end{equation}
and the Hamiltonian in terms of the oscillators is
\begin{equation}
{\cal H}_{b} ={{1}\over{2}}\sum_{n=0}^{\infty}\omega_n\,
(a_{n}^{i}a_{n}^{i\dagger} + a_{n}^{i\dagger}a_{n}^{i}).
\end{equation}
For the fermionic mode expansions, we note that $M$ is the product of
$\gamma^i$'s. Since $S_L$ and $S_R$ are of the same chirality the
number of $\gamma^i$'s has to be even. We define sign factors
$\eta,\xi^i=\pm 1$ as follows,
\begin{equation}
\gamma^i M=\xi^i M \gamma^i, \quad M^T=\eta M,
\end{equation} 
and decompose M into $4\times 4$ matrices:
\begin{equation}
MS_{R,L}=\left(\begin{array}{c}Ns_{R,L}\\
\hat{N}\hat{s}_{R,L}\end{array}\right).
\end{equation}
The solution for the D9-brane requires $M=1,\eta=1,\xi_{1}\xi_2=1$ and
D5-brane along $(+,-,1,2,3,4)$ directions has $M = \gamma^{1234}, \eta
= 1,\xi_{1}\xi_2=1$. We are interested in only these two cases for the
purpose of this paper. We quote the fermionic mode expansions relevant
to this condition\cite{yoji}.
\begin{eqnarray} 
s_L&=&\frac{\sqrt{\alpha'}}{2}{\Bigg
[}\sqrt{2}(\cos{m\tau}+\sin{m\tau}\,\Lambda)s_0 \\ \nonumber
&+&\sum_{n=-\infty,n\neq
0}^{\infty}e^{-i\omega_n\tau}\left(\sqrt{\frac{\omega_n+n}{\omega_n}}
e^{-in\sigma}-{\rm
sgn}(n)i\sqrt{\frac{\omega_n-n}{\omega_n}}e^{in\sigma}\Lambda\right)
s_n{\Bigg ]}, \cr
s_R&=&\frac{\sqrt{\alpha'}}{2}{\Bigg
[}\sqrt{2}(\sin{m\tau}\Lambda+\cos{m\tau})s_0\\ \nonumber
&+&\sum_{n=-\infty,n\neq 0}^{\infty}e^{i\omega_n\tau}
\left(\sqrt{\frac{\omega_n+n}{\omega_n}}e^{in\sigma}
- {\rm sgn}(n)i\sqrt{\frac{\omega_n-n}{\omega_n}}
e^{-in\sigma}\Lambda\right)s_n{\Bigg ]} ,
\end{eqnarray}
with the anti-commutation relations
\begin{equation}
\{s_n^{\alpha},s_{n'}^{\beta}\}=\delta^{\alpha,\beta}\delta_{n,n'},
\quad s_n^{\dagger}=s_{-n},
\end{equation}
with the Hamiltonian given by
\begin{equation}
{\cal H}_{f} = -{i\over 2}m s_0\Lambda s_0+{1\over2}
\sum_{n=-\infty,n\neq 0}^{\infty}\omega_n s_{-n}s_n.
\end{equation}
Along with these we further have four massless scalars and four
massless fermions as in the case of closed string. We again avoid
writing them down here. The total light-cone Hamiltonian is 
\begin{equation}
\label{openlcham}
H_{l.c} = -2p_+ = {1\over p^+}\big({p_\alpha^2 + {\cal H}_{o}}\big), 
~~~ {\cal H}_{o} = {\cal H}_b + {\cal H}_f + \cdots
\end{equation}
where the dots represent the contributions from the four massless
bosons and fermions.

\section{\bf Regularizing the Zero Point Energies}
In this appendix we use $\zeta$ function regularization for the normal
ordering constants of PB, PF, AB and AF cases.
 
We need the zero-point energies for periodic bosons, periodic fermions
and anti-periodic fermions in our calculation. Let us start with the
periodic bosons. From the Hamiltonian for open strings, we see that
the normal ordering constant is given by
\begin{equation}
\frac{1}{4}\sum_{n=-\infty}^{\infty}\omega_{n}=\frac{1}{4}
\sum_{n=-\infty}^{\infty}\sqrt{n^2+m^2}
\end{equation}
We regularize the right-hand-side as follows. We follow the method
used for example in \cite{cftbook}.
\begin{eqnarray}
\label{proof1}
&&\sum_{n=-\infty}^{\infty}\sqrt{n^2+m^2}=\sum_{n}\frac{1}{|n+im|^{-1}}
=\sum_{p\in {\mathbb Z}}\int_{0}^{1}dy\, e^{-2\pi
ipy}\sum_{n\in {\mathbb Z}}\frac{1}{[(n+y)^2+m^2]^{-{1\over 2}}} \cr
&&= \sum_{p\in {\mathbb Z}}\sum_{n}\int_{n}^{n+1}dy\,e^{-2\pi i
py}\frac{1}{(y^2+m^2)^{-{1\over 2}}}
= \sum_{p\in {\mathbb Z}}\int_{-\infty}^{\infty}dy\,e^{-2\pi i
py}\frac{1}{(y^2+m^2)^{-{1\over 2}}}\\
&&=\frac{\sqrt{\pi}}{\Gamma [-{1\over 2}]}\sum_{p\in {\mathbb Z}}
\int_{-\infty}^{\infty}dy\,e^{-2\pi i
py}\int_{0}^{\infty}dt\,t^{-{3\over 2}}e^{-t(y^2+m^2)} 
=\frac{\sqrt{\pi}}{\Gamma [{-{1\over 2}}]}\sum_{p\in {\mathbb Z}}
\int_{0}^{\infty} dt\,
t^{-2}e^{-\frac{\pi^2 p^2}{t}-tm^2}\nonumber
\end{eqnarray}
Now performing a change of variables $s=\pi^2/t$, we get the result as
\begin{equation}
-{1\over 2}\frac{1}{(2\pi)^{2}}\sum_{p\in {\mathbb Z}}\int_{0}^{\infty}ds\,
e^{-p^2s-\frac{m^2\pi^2}{s}},
\end{equation}
where we have used $\Gamma[{-{1\over 2}}]=-2\sqrt{\pi}$.
Regularizing the above result by removing the infinite $p=0$ part
gives
\begin{equation}
-\frac{1}{(2\pi)^{2}}\sum_{p=1}^{\infty}\int_{0}^{\infty}ds\,
e^{-p^2s-\frac{m^2\pi^2}{s}}
\end{equation}
This result is exactly the same as given in \cite{bgg}. The fermionic
result is negative of the above as expected from space-time
supersymmetry.

The case for anti-periodic fermions can be analyzed using the above
approach. In that case we need to replace $n$ by $n-\frac{1}{2}$ as a
result of which we will get an additional factor of $(-1)^p$ equation
(\ref{proof1}) and in the final answer. Again this matches exactly
with the result quoted in \cite{bgg} (see also
\cite{PandoZayas}).{\footnote{We thank Yuji Sugawara for an e-mail
correspondence on this point.}}

\end{document}